\documentclass[]{an}
\usepackage{graphicx,times,fancyhdr}
\usepackage{afterpage}
\usepackage{amsmath}
\usepackage{graphicx}

\begin{document}

\def\apj{{ApJ}}       
\def\mn{{MNRAS}} 
\def\aa{{A\&A}} 
\def\qq{\qquad\qquad}                      
\def\qqq{\qquad\qquad\qquad}               
\def\q{\qquad}

\renewcommand{\vec}[1]{\mbox{\boldmath $#1$}}
\def\bl{\par\vskip 12pt\noindent}

\def\Pm{\mathop{\rm Pm}\nolimits}
\def\bib{\item}
\def\beg{\begin{eqnarray}}
\def\ende{\end{eqnarray}}
\def\Om{{\it \Omega}}
\def\Del{{\it \Delta}}
\def\Gam{{\it \Gamma}}
\def\gsim{\lower.4ex\hbox{$\;\buildrel >\over{\scriptstyle\sim}\;$}}
\def\lsim{\lower.4ex\hbox{$\;\buildrel <\over{\scriptstyle\sim}\;$}}

\title{Do mean-field dynamos in nonrotating   turbulent shear-flows exist?}
\author{G. R\"udiger\inst{1} \and L.L. Kitchatinov\inst{1,2}}
\institute{Astrophysikalisches Institut Potsdam,
         An der Sternwarte 16, 14482 Potsdam, Germany 
\and
Institute for Solar-Terrestrial Physics, P.O. Box 4026, 664033,
Irkutsk, Russia 
}

\date{Received  2006; accepted  2006; published online 2006}

\abstract{A plane-shear flow in a fluid with  forced turbulence is considered. 
If the fluid is electrically-conducting then a  mean
electromotive force (EMF) results even without  basic rotation and  the magnetic  diffusivity
becomes a highly anisotropic  tensor. It is checked whether in this case   self-excitation of a large-scale 
magnetic field is possible (so-called  $\bar{\vec{W}}\times \bar{\vec{J}}$-dynamo) and the answer is  NO.  The calculations reveal the cross-stream
components of the  EMF perpendicular to the mean
current having the wrong signs, at least for small magnetic Prandtl numbers. After our results  numerical simulations with magnetic Prandtl number $\simeq 1$   have  only a  restricted meaning as the Prandtl number dependence of the diffusivity tensor  is rather strong. 
\\
   If, on the other hand, the turbulence field is
stratified in the vertical direction  then a dynamo-active $\alpha$-effect is produced.
The critical magnetic Reynolds number for such a self-excitation
in a simple shear flow is slightly above 10 like for the  other
-- but much more complicated -- flow patterns used in existing
dynamo experiments with liquid sodium or gallium.
}


\correspondence{gruediger@aip.de}

\maketitle

\section{Introduction}
One of the most challenging problems in astrophysical fluid
dynamics is the investigation of the interaction of rotation
and/or magnetic fields with turbulence. The maintenance of
differential rotation and the induction of global magnetic fields
can be the consequences of such interactions. In most of the
cases, however, the turbulence must be anisotropic and/or
inhomogeneous to generate  interesting global phenomena. There is
one exception, however, where homogeneous and isotropic
turbulence interacting with an inhomogeneous magnetic field leads
to the well-known $\vec{\Om} \times \vec{J}$-term  in the
turbulent electromotive force (EMF) which together with
differential rotation can  be dynamo-active  (Krause \& R\"adler 1980). The term, however, 
vanishes for spectra of the mixing-length type (Kitchatinov, Pipin \& R\"udiger 1994).

Rotation may not be the only flow whose influence enables the
turbulence to generate global fields. Any vortical large-scale
motion can be suspected to do the same. The simplest case to
probe  the expectation theoretically or in  laboratories is,
probably, the plane shear flow.

In the present paper  a plane shear flow is considered  to
ana\-lyze the main phenomena of the mean-field
magnetohydrodynamics  and to formulate  predictions for its
experimental realization. We shall show that the interaction of
free homogeneous and isotropic turbulence with a plane shear flow
does {\em not} lead to  the so-called
$\bar{\vec{W}}\times \bar{\vec{J}}$-dynamos (cf. Rogashevskii \& Kleeorin 2003). 
Large-scale dynamos are possible only if the
turbulence is not uniform along a direction normal to the plane
of the shear. In this case an $\alpha$-effect is produced by the
sheared turbulence which together with the shear itself generates
global magnetic fields.
\section{EMF of sheared turbulence}
The magnetic-diffusivity tensor $\eta_{ijk}$ relates the mean electromotive force (EMF)
 \beg
  {\vec{\cal E}}= \overline{\vec{u}' \times \vec{B}'}
  \label{26}
 \ende
to the gradients of the mean magnetic field via the relation
 \beg
  {\cal E}_i=\eta_{ijk} \bar B_{j,k}.
  \label{27}
 \ende
If the influence of the shear flow is included to  first order
the general structure of the diffusivity tensor is
 \begin{eqnarray}
  && \eta_{ijk}=\eta_0 \epsilon_{ijk}+\eta_1 \epsilon_{ijp} \bar
  u_{k,p}+\eta_2 \epsilon_{ijp} \bar u_{p,k}+
  \nonumber\\
  && \quad\quad\quad + \eta_3 \epsilon_{ikp} \bar u_{j,p} + \eta_4
  \epsilon_{ikp} \bar u_{p,j}.
  \label{28}
 \end{eqnarray}
Quasilinear derivations of the coefficients in (\ref{28}) can now be performed.
The electromotive force (\ref{26}) may be constructed by a perturbation method.
The fluctuating  fields are represented by a series expansion,
\beg
  {\vec F}' = {\vec F}^{(0)} + {\vec F}^{(1)} + {\vec F}^{(2)} + ...\ ,
  \label{7}
\ende
where the upper index shows the order of the  contributions  in terms of the mean shear flow and magnetic field.

The zero-order terms represent the \lq original' isotropic turbulence not yet
influenced by the shear. The spectral tensor for the original turbulence is
 \beg
 \hat Q^{(0)}_{ij}=\frac{E(k,\omega)}{16\pi k^2} \left(\delta_{ij}-\frac{k_i
 k_j}{k^2}\right),
 \label{8}
 \ende
where the positive-definite spectrum $E$ gives the intensity of isotropic
fluctuations, i.e.
 \beg
 \overline{{\vec{u}^{(0)}}^2}=\int\limits_0^\infty\!\!\int\limits_0^\infty E(k,\omega) \
 {\rm d}k \ {\rm d}\omega .
 \label{9}
 \ende
We apply the quasilinear approximation (SOCA) to derive the
higher-order terms in (\ref{26}). They are generally found by a perturbation
method from the linearized equations. E.g. the linearized momentum equation reads
 \beg
 &&\frac{\partial u^{(n)}_i}{\partial t} - \nu \Delta u^{(n)}_i =
 \nonumber \\
 &&\ \ \ \ \ - \frac{\partial}{\partial
 x_k}\left(\bar u_k u^{(n-1)}_i+\bar u_i  u^{(n-1)}_k\right) - \frac{\partial p^{(n)}}{\partial
 x_i} ,
 \label{10}
 \ende
where  the upper index shows the order in the mean shear. With this
equation taken for $n=1$, the first-order correction, ${\vec u}^{(1)}$, can be
expressed in terms of the given original turbulence. 

As known, for the isotropic eddy diffusivity $\eta_0$ one finds
 \beg
  \eta_0=\frac{1}{3} \int\limits_0^\infty\!\!\int\limits_0^\infty
  \frac{\eta k^2 E(k,\omega)}{\omega^2+\eta^2 k^4}
  {\rm d}k \ {\rm d}\omega .
  \label{29}
 \ende
(see R\"udiger \& Hollerbach 2004). The shear-related coefficients of (\ref{28}) read
 \beg
  \eta_i = \int\limits_0^\infty\!\!\int\limits_0^\infty
  C_i(k,\omega) E(k,\omega){\rm d}k \ {\rm d}\omega ,
  \label{30}
 \ende
for $i=1\dots 4$ with the nontrivial kernels
 \beg
 && C_1 = -\frac{5\nu^3\eta^3k^{12} + \omega^2\nu\eta
  \left( \eta^2 + 3\nu^2 + 6\nu\eta\right)k^8}
  {15 \left(\omega^2+\nu^2k^4\right)^2\left(\omega^2 +
  \eta^2k^4\right)^2}-
  \nonumber \\
 && - \frac{\omega^4\eta \left( 6\eta - \nu\right)k^4}
  {15\left(\omega^2+\nu^2k^4\right)^2\left(\omega^2 +
  \eta^2k^4\right)^2}
   - \frac{\eta^2k^4\left(\eta^2k^4 - 3\omega^2\right)}
  {15\left(\omega^2 + \eta^2k^4\right)^3} ,
  \nonumber \\
 && C_2 = -\frac{5\nu^3\eta^3k^{12} + \omega^2\nu\eta
  \left( \eta^2 + 3\nu^2 - 6\nu\eta\right)k^8}
  {15 \left(\omega^2+\nu^2k^4\right)^2\left(\omega^2 +
  \eta^2k^4\right)^2}+
  \nonumber \\
 && + \frac{\omega^4\eta \left( 6\eta + \nu\right)k^4}
  {15 \left(\omega^2+\nu^2k^4\right)^2\left(\omega^2 +
  \eta^2k^4\right)^2}
   + \frac{\eta^2k^4\left(\eta^2k^4 - 3\omega^2\right)}
  {15\left(\omega^2 + \eta^2k^4\right)^3} ,
  \nonumber \\
 && C_3 =
  -\frac{8 \omega^2\eta^2 k^4 }
  {15 \left(\omega^2+\nu^2k^4\right)\left(\omega^2 +
  \eta^2k^4\right)^2} ,
  \nonumber \\
 && C_4 =
  \frac{4 \omega^2\eta^2 k^4 }
  {15 \left(\omega^2+\nu^2k^4\right)\left(\omega^2 +
  \eta^2k^4\right)^2}+\nonumber\\
  && \quad\quad\quad\quad\quad\quad + \frac{5\omega^4 - 6\omega^2\eta^2k^4 - 3\eta^4k^8}
  {15\left(\omega^2 + \eta^2k^4\right)^3} .
  \label{31}
 \ende
$C_3$ is negative-definite. Another  important  property  here is that the kernels $C_3$ and $C_4$ also exist  in the limit $\nu \to 0$, i.e.
\beg
&& C_3= - \frac{8}{15} \frac{\eta^2 k^4}{(\omega^2 + \eta^2 k^4)^2},
\nonumber\\
&& C_4= \frac{1}{15} \frac{5\omega^4 -2\omega^2 \eta^2 k^4+ \eta^4 k^8}{(\omega^2 +\eta^2 k^4)^3}.
\label{c3c4}
\ende
For sufficiently small magnetic Prandtl number ${\rm Pm}=\nu/\eta$ the kernel $C_4$ is thus positive-definite. This is not true, however, for Pm of the order unity. For ${\rm Pm}=1$ one obtains from (\ref{31})$_4$
\beg
C_4= \frac{1}{15} \frac{(\omega^2-\eta^2 k^4)(5\omega^2 + 3\eta^2 k^4)}{(\omega^2 +\eta^2 k^4)^3},
\label{c4}
\ende
which  has no definite sign. The high-frequency parts of the spectrum provide positive contributions to the integral and the low-frequency parts provide negative contributions to the integrals. White noise (with $\partial E/\partial \omega=0$) leads to positive-definite values of $\eta_4$ while the $\tau$-approximation ($E\propto \delta(\omega)$) leads to $\eta_4<0$\footnote{in this case  $\eta_3=0$ and no slab dynamo is possible, see Eq. (\ref{W})}. Numerical simulations with ${\rm Pm}\simeq 1$  do have thus {\em only a  restricted meaning} as the Prandtl number dependence of $\eta_4$ is rather strong. 

The so-called $\tau$-approximation can be applied as a crude representation of
nonlinear turbulent effects. The approximation replaces the left side of
Eq.~(\ref{10}) by  $u^{(n)}_i/\tau_{\rm
corr}$, where $\tau_{\rm corr}  = \ell_{\rm corr}/u'$ is the eddy turnover
time ($\ell_{\rm corr}$ is the correlation length). The
$\tau$-approximation can simply be recovered by substituting $\omega = 0$
and $\nu k^2 = \tau_{\rm corr}^{-1}$. In particular, the spectrum
\beg
  E(k,\omega ) = 2\overline{u'^2}\delta (\omega )\delta (k - \ell_{\rm corr}^{-1}),
     \label{15}
\ende
with $\ell_{\rm corr}^2/\nu = \tau_{\rm corr}$ can be used to derive well-established estimates of the turbulence-induced coefficients (\ref{30}) (Vainshtein 1980; Vainshtein \& Kitchatinov  1983).

Only $\eta_3$ and $\eta_4$ are important for the simple slab dynamo model discussed below. 
Consider a shear flow with  uniform vorticity in the vertical $z$-direction,
i.e.
\beg
\bar u_y=S x.
\label{1}
\ende
The shear flow may exist in a turbulence field which does not
possess any other anisotropy apart from that induced by the shear
(\ref{1}) itself. 
For experiments in the laboratory it might be relevant that the relations
\beg
{\cal E}_x= \eta_0 D \bar B_y - \eta_4 S D \bar B_x
\ende
and 
\beg
{\cal E}_y= -\eta_0 D \bar B_x + \eta_3 S D \bar B_y
 \label{exey}
 \ende
($D={\rm d}/{\rm d}z$) follow from (\ref{27}) and (\ref{28}) for a $z$-dependent field
imposed in horizontal direction. If  the field is imposed in $x$-direction, we have
\beg
 {\cal E}_x= - \eta_4 S D \bar B_x,
  \label{32}
\ende
and if the field is imposed in the $y$-direction then
\beg
  {\cal E}_y=  \eta_3 S D \bar B_y.
  \label{calEy}
\ende
 The sign of $\eta_4$ is thus opposite
to the sign of the expression ${\cal E}_x S \bar B_{x,z}$ and the sign of $\eta_3$ is the same as the sign of ${\cal E}_y S \bar B_{y,z}$. Note
that the EMF components are  {\em perpendicular} to the
mean current $\bar{\vec{J}}$.  The standard diffusion-induced EMF
(without shear)  is always parallel or antiparallel to  the mean current. 
The EMF due to the standard $\alpha$-effect is also
{\em parallel} to the mean magnetic field.

The  quasilinear theory provides a positive $\eta_4$ 
 for  small Pm  and a negative $\eta_3$ in all cases. We shall see below that this constellation 
does {\em not} allow a simple `$\bar{\vec{W}}\times \bar{\vec{J}}$'
slab dynamo. Our negative result mainly bases on the positive sign of
$\eta_4$. Whether $\eta_4$ is indeed positive must be checked with
laboratory experiments and/or with numerical simulations for various magnetic Prandtl numbers 
(Brandenburg 2005).
\subsection{Alpha effect}
Now the nondiffusive part of the mean EMF (\ref{26}) is
considered. If an $\alpha$-effect exists in the shear flow we have
 \beg
  {\cal E}_i=\alpha_{ij} \bar B_j + \dots .
  \label{33}
 \ende
(for more details see R\"udiger \& Hollerbach 2004). The tensor $\alpha$ must be a pseudotensor so that
an $\epsilon$-tensor has to appear in the $\alpha$-coefficients. The
construction of the EMF  ${\cal E}_i =
\epsilon_{ijk}\overline{u'_j B'_k}$ is the only possibility for
the $\epsilon$-tensor to appear. Therefore, the subscript of
${\cal E}_i$ is always a subscript of the $\epsilon$-tensor. As
the $\epsilon$-tensor is of 3$^{\rm rd}$ rank an inhomogeneity of
turbulence with the stratification vector, $\vec{g}=\nabla \log
\overline{u'^2}$, must also be present for the $\alpha$-effect to
exist. If the shear flow is included to its first order, the
general structure of the $\alpha$-tensor is
 \beg
  \alpha_{ij} &=&
  \gamma \epsilon_{ijk}  g_k +
  \big(\alpha_1 \epsilon_{ikl} \bar{u}_{j,k} +
 \alpha_2 \epsilon_{ikl} \bar u_{k,j} \big) g_l
 +\nonumber\\
 && \ \ \ +\alpha_3 \epsilon_{ikl} g_j  \bar u_{l,k}
 + \alpha_4 \epsilon_{ikj} \bar u_{l,k} g_l
 + \alpha_5 \epsilon_{ijk} \bar u_{k,l} g_l .
 \label{34}
 \ende
If the stratification is along the vertical $z$-axis it follows from (\ref{34})  that
 \beg
  \alpha_{xx}= \alpha_2 g_z S= \alpha_x S, \ \ \ \ \ \ \ \ \ \ \alpha_{yy}= -\alpha_1 g_z S= \alpha_y S,
\ende
and 
\beg
  \alpha_{xy}=-\alpha_{yx}   =\gamma g_z =\Gamma .
 \label{35}
 \ende
The anisotropy of the $\alpha$-tensor is described by the
difference between $\alpha_x$ and $\alpha_y$. The so-called turbulence-induced diamagnetism is described by $\alpha_{xy}$ (see Krause \& R\"adler 1980). The coefficients of (\ref{34}) read
 \beg
  \gamma = \frac{1}{6}\int\limits_0^\infty\!\!\int\limits_0^\infty
  \frac{\eta k^2 E(k,\omega)}{\omega^2+\eta^2 k^4}
   {\rm d}k \ {\rm d}\omega
   \label{36}
 \ende
for the pumping term and
 \beg
  \alpha_i = \int\limits_0^\infty\!\!\int\limits_0^\infty
  A_i E\left( k,\omega \right) {\rm d}k \ {\rm d}\omega ,\ \ \
  i = 1\dots 5,
  \label{37}
 \ende
for the $\alpha$-effect with
 \beg
  A_1 &=& \frac{4\nu\eta^3k^8 + 2\omega^2\eta \left(\nu + \eta\right)k^4}
  {15 \left(\omega^2+\nu^2k^4\right)\left(\omega^2 +
  \eta^2k^4\right)^2} +
  \nonumber \\
  && \ \ \ \ \ \ \ \ +\frac{\eta^2k^4\left(\eta^2k^4 - 3\omega^2\right)}
  {15\left(\omega^2 + \eta^2k^4\right)^3} ,
  \nonumber \\
  A_2 &=&  -\frac{\eta^2\nu^3\left(4\eta-5\nu\right)k^{12}}
  {60 \left(\omega^2+\nu^2k^4\right)^2\left(\omega^2 +
  \eta^2k^4\right)^2}-
  \nonumber\\
  &&\ \ \ -\frac{\omega^2\nu\left(28\eta^3 - 4\eta^2\nu + 12\eta\nu^2
  + 5\nu^3\right)k^8}
  {60 \left(\omega^2+\nu^2k^4\right)^2\left(\omega^2 +
  \eta^2k^4\right)^2} -
  \nonumber\\
  &&\ \ \ -\frac{\omega^4\eta\left(\eta + 36\nu\right)k^4 - 5\omega^6}
  {60 \left(\omega^2+\nu^2k^4\right)^2\left(\omega^2 +
  \eta^2k^4\right)^2}
  \label{38}
 \ende
for two of the kernels. Only the terms occurring in (\ref{35})
have been given. For small Pm, the  $\alpha_x$ and $\alpha_y$
are  both of the
same sign which is opposite to the sign of $g_z$. The $\alpha_x$
appears to be smaller than the $\alpha_y$.
\section{Slab dynamos}
The shear-flow dynamos without and with $\alpha$-effect are
studied with a 1D slab model. The dynamo region is finite in the
$z$-direction but unlimited in $x$ and $y$. The equations for the
mean magnetic field with zero $\bar{B}_z$-component read
  \beg
  && \frac{\partial \bar B_x}{\partial t}
  - \eta_0 D^2 \bar B_x = \Gamma
  D\bar B_x -\alpha_y  S D \bar B_y-\eta_y SD^2 \bar B_y,
  \nonumber\\
  && \frac{\partial \bar B_y}{\partial t}
  - \eta_0 D^2 \bar B_y =
  \nonumber\\
  && \quad\quad\quad\quad \Gamma D\bar B_y + \alpha_x
  S D \bar B_x + S \bar B_x-\eta_x SD^2 \bar B_x
  \label{39}
 \ende
with  $D = {\rm d}/{\rm d}z$ and
 \beg
  \eta_x=\eta_4 , \ \ \ \ \ \quad\quad \eta_y= \eta_3.
  \label{40}
 \ende
The equations can be normalized with
 \beg
  \hat \alpha = \frac{\alpha}{H}, \
  \hat\Gamma = \frac{H\Gamma}{\eta_0}, \
  \hat S=\frac{H^2 S}{\eta_0} , \
  \hat\eta_{x}=\frac{\eta_{x}}{H^2},\
  \hat\eta_{y}=\frac{\eta_{y}}{H^2},
  \label{41}
 \ende
so that
 \beg
   \frac{\partial \bar B_x}{\partial t}
   -\frac{{\rm d}^2 \bar B_x}{{\rm d} z^2}
   &=& \hat \Gamma \frac{{\rm d} \bar B_x}{{\rm d}z}
   - \hat\alpha_y \hat S \frac{{\rm d} \bar B_y}
   {{\rm d}z}- \hat\eta_y \hat S
   \frac{{\rm d}^2\bar B_y}{{\rm d}z^2} ,
   \nonumber\\
   \frac{\partial \bar B_y}{\partial t}-\frac{{\rm d}^2
   \bar B_y}{{\rm d} z^2}&=&
   \hat \Gamma \frac{{\rm d} \bar B_y}{{\rm d}z} +
   \left(\hat\alpha_x \frac{{\rm d}  \bar B_x}
   {{\rm d}z} + \bar B_x\right) \hat S-
   \nonumber\\
   &-&\hat\eta_x \hat S
   \frac{{\rm d}^2 \bar B_x}{{\rm d}z^2}
   \label{42}
 \ende
results. The vacuum boundary conditions
$
   \bar B_x(0)=\bar B_y(0)= \bar B_x(1)=\bar B_y(1)=0
$
are applied.
 \begin{figure}[htb]
   \centering
   \includegraphics[width=7cm]{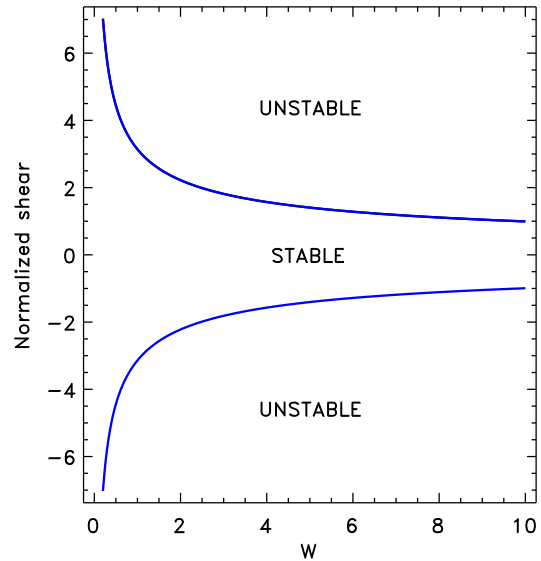}
      \caption{Map for the  dynamo instability produced by the
               $\bar{\vec{W}}\times \bar{\vec{J}}$-effect. Field excitation can be
               found for sufficiently large positive values of the
               $W$-parameter (\ref{W}).}
   \label{f1}
 \end{figure}
\subsection{Dynamos without stratification}
Let  the preferred direction $\vec{g}$  be absent in  a uniform
original turbulence. Alpha-effect and pumping of magnetic field
both vanish in this case. Nevertheless, a hydromagnetic dynamo by
the  $\bar{\vec{W}}\times \bar{\vec{J}}$-effect might  be
expected. Figure~\ref{f1} shows the corresponding stability map. Dynamo  excitation  indeed exists for 
 \beg
 W = \hat\eta_y\left( 1 + \pi^2\hat\eta_x\right)>0.
 \label{W}
 \ende
Note that after (\ref{31})$_3$  $\eta_y=\eta_3$ vanishes in the $\tau$-approximation or is otherwise  {\em negative-definite}. Plane-shear flow dynamos are thus not possible  in $\tau$-approximation -- and they are also not possible without inclusion of $\eta_x$. Only the product $\eta_x \eta_y$ together with the overall shear allows the existence of  plane-shear flow dynamos. No dynamo linear in the $\vec{W}\times \vec{J}$ term can exist in plane geometry. 

For small Pm, however,  $\eta_x$ is positive after Eq.~(\ref{c3c4})$_2$.
Therefore also for the plane-shear flow (\ref{1}) of
liquid sodium or gallium ($\mathrm{Pm} \lsim 10^{-5}$) a
$\bar{\vec{W}}\times \bar{ \vec{J}}$-slab-dynamo mechanism proves
to be impossible. It would thus be  challenging to probe  the
sign of the $\eta_x$ in the laboratory  in accordance with the Eq.~(\ref{32}).

Another situation holds for the case of magnetic Prandtl number of order unity which is  
used in most of the  numerical MHD simulations. As outlined above, the $\eta_4$ can then be negative  
if the frequency spectrum of the turbulence is steep enough (if it is too steep 
then $\eta_3$ vanishes!). 
A limiting magnetic Prandtl number between the two cases must exist and can be fixed by numerical integrations.
 \begin{figure}[htb]
   \centering
   \includegraphics[width=8cm]{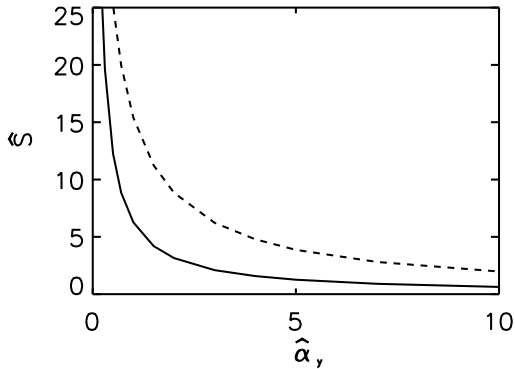}
   \caption{Stability map for a global dynamo solution of
            Eqs.~(\ref{42}) for $\hat\Gamma =-0.5$.
        Magnetic fields are excited in the parameters region
        above the lines.
        Solid: $\hat \alpha_x= \hat \alpha_y$,
        dashed: $\hat \alpha_x= 0.1 \hat \alpha_y$.}
   \label{slab2}
 \end{figure}
\subsection{Dynamos with stratification}
The $\alpha$-effect dynamos may  be considered now ignoring  the
anisotropic dissipation in the Eqs.~(\ref{42}). For given shear,
self-excitation of magnetic fields can be observed for
sufficiently large $\hat\alpha_y$ and/or large enough shear. All
the solutions are oscillatory. The oscillation time is of the
order of the diffusion time for the case of isotropic
$\alpha$-effect. In both considered cases a typical condition for
dynamo excitation is
 \beg
  \hat\alpha_y \hat S > 10.
  \label{44}
 \ende
(Fig.~\ref{slab2}).
A reformulation of this relation with $S\simeq\bar u/L$ yields
\beg
{\rm Rm} > 10 \frac{L^2}{\alpha_y H},
 \label{45}
\ende
where Rm=$L\bar u/\eta_0$ is the magnetic Reynolds number of the mean flow defined
with the eddy diffusivity (\ref{29}). The estimates $\eta_0
\simeq u' \ell_{\rm corr}/3$ and $\alpha_y \simeq \ell_{\rm
corr}^2/H$ then reduce the inequality (\ref{45}) to the condition
 \beg
   u' < \frac{\bar u \ell_{\rm corr}}{3 L} .
   \label{46}
 \ende
The fluctuating velocity should be sufficiently small to produce
a dynamo. It cannot be too small, however, as the eddy
diffusivity must remain larger than the microscopic one, $\eta_0
> \eta$. This finally yields the condition
 \beg
   \frac{\eta}{\ell_{\rm corr}} < u' < \frac{\bar u \ell_{\rm corr}}{3 L},
   \label{47}
 \ende
which must be fulfilled for dynamo self-excitation. The
microscopic magnetic diffusivity $\eta$ of liquid sodium or
gallium is about $0.1$~m$^2$/s. If the channel width, $L$, is  as
large as 1~m, then for the most optimistic case ($\ell_{\rm corr}
\sim L$) the range (\ref{47}) can be realized with shear
velocities exceeding 1~m/s which is of the same order as in other
dynamo experiments with  a much more complicated geometry.
\section{Summary}
It is suggested that the basic effects of the theory of the
turbulent dynamo which are usually concerned as special
properties of rotating fluids can also be found for the plane
shear flow. The same is probably true for any flow with global
vorticity. The elementary structure of the shear flow largely
simplifies the consideration. In the Appendix we have shown with the same concept as above 
that the generation of large-scale vorticity by a uniformly
sheared turbulence is not possible and the generation of magnetic
fields by such a flow  can only be hoped for large magnetic
Prandtl numbers.

The dynamo instability can be realized, however, when the
turbulence is not uniform. Then the  stratified turbulence
produces the  $\alpha$-effect which  can excite an
oscillatory mean magnetic field in the shear flow. This opens a
possibility for the realization of a turbulent dynamo in the
laboratory with a quite simple flow geometry -- if a nonuniform
turbulence can be created in the  channel flow. Estimates of
the excitation condition for such a dynamo, however, shows   it  to be  at the
limit of  current experimental possibilities.
\begin{acknowledgements}
L.L.K. is grateful to the Alexander von Humboldt Foundation and to Astrophysical Institute Potsdam for hospitality and the visitors support. Thanks are also due to the Russian Foundation for Basic Research (Project 05-02-16326).
\end{acknowledgements}

\appendix
\section{Hydrodynamic stability of the shear flow}
It is important for the above consideration that the  linear shear flow   (\ref{1}) is hydrodynamically stable under the
presence of  {\em nonlinear} shear  terms in the correlation tensor $Q_{ij}$. This question has
been addressed by   Elperin, Kleeorin and Rogashevskii (2003). With a
dispersion relation formulated on the basis of (\ref{5}, below) an 
instability has been constructed for plane wave disturbances with
spatial inhomogeneity in the vertical $z$-direction. In contrast to this, we shall show that within the first-order smoothing approximation the shear flow is stable.

The one-point correlation tensor
 \beg
 Q_{ij}=\overline{u'_i(\vec{x},t) u'_j(\vec{x},t)} \label{2}
 \ende
in its linear form  reads
 \beg
 Q_{ij}= P \delta_{ij} - \nu_0 \left(\bar u_{i,j}+ \bar u_{j,i}\right).
 \label{3}
 \ende
Here $\nu_0$ is the isotropic eddy viscosity; the turbulent
pressure, $P$, includes all coefficients of the Kronecker tensor
$\delta_{ij}$.

The experiment by Champagne et al. (1970) with
sheared turbulence indicates that the linear relation (\ref{3})
cannot be the whole truth. In the experiment the rms
downstream velocities are systematically larger than the
rms velocities in the cross-stream  direction. The turbulence
intensities for these two directions should, however, be equal
after Eq.~(\ref{3}) which can be read as
 \beg 
\overline{u'^2_y}=
 P, \quad \overline{u'^2_x}=P, \quad \overline{u'_x u'_y}=-\nu_0 S.
 \label{4}
 \ende
The same remains  true if higher-order derivatives such as $\bar
u_{i,jll}+\bar u_{j,ill}$ are included.

However, if the mean shear is indeed the only reason for
anisotropy, one has also to involve  nonlinear terms, i.e.
 \beg
 Q_{ij}&=& P \delta_{ij} - \nu_0\left(\bar u_{i,j}+\bar u_{j,i}\right)
 + \nu_1 \bar u_{i,k} \bar u_{j,k}+
 \nonumber \\
  && \ \ \ \ \ \ \ +\nu_2 \bar u_{k,i} \bar u_{k,j}
  + \nu_3\left(\bar u_{i,k} \bar u_{k,j} + \bar u_{j,k} \bar u_{k,i}\right).
 \label{5}
 \ende
By this expression the horizontal intensities can differ if the
coefficients $\nu_1$ and $\nu_2$ of the nonlinear terms do not
coincide,
 \beg
 \overline{u'^2_y}-\overline{u'^2_x}= (\nu_1-\nu_2) S^2.
 \label{6}
 \ende
It should be $\nu_1 > \nu_2$ for agreement with the aforementioned shear-flow experiment.

The first-order term of Eq.~(\ref{3}) is  the eddy viscosity
 \beg
 \nu_0=\frac{4}{15} \int\limits_0^\infty\!\!\int\limits_0^\infty \frac{\nu^3 k^6 E(k,\omega)}{(\omega^2+\nu^2
 k^4)^2} {\rm d}k\ {\rm d}\omega,
 \label{11}
 \ende
(Krause \& R\"udiger 1974). 
The second-order correction to the correlation tensor 
reproduce the nonlinear  terms of (\ref{5}). The coefficients
$\nu_1$, $\nu_2$ and $\nu_3$ result as
 \beg
 \nu_i = \int\limits_{0}^{\infty}\!\!\int\limits_0^\infty
  K_i(k,\omega) E(k,\omega){\rm d}k\ {\rm d}\omega , \ \ \ n=1\dots 3\ ,
 \label{13}
 \ende
with the kernels
 \beg
 &&  K_1 = \frac{25\nu^6k^{12} + 63\nu^4k^8\omega^2 - 149\nu^2k^4\omega^4
       + 5\omega^6}{105 \left(\omega^2 + \nu^2k^4\right)^4},
 \nonumber \\
 &&  K_2 = \frac{4 \left(8\nu^6k^{12} + 7\nu^4k^8\omega^2 -
       46\nu^2k^4\omega^4
       + 3\omega^6\right)}{105 \left(\omega^2 + \nu^2k^4\right)^4},
 \nonumber \\
 &&   K_3 = \frac{25\nu^6k^{12} + 49\nu^4k^8\omega^2 - 149\nu^2k^4\omega^4
       + 19\omega^6}{105 \left(\omega^2 + \nu^2k^4\right)^4}.
 \label{14}
 \ende
All the kernels have negative (third) terms in their numerators.
Nevertheless, all the coefficients (\ref{13}) are \lq almost
always' positive. They are positive definite in the most
popular simplifying cases, i.e.

-- within the $\tau$-approximation: 
\beg
 \nu_1 = \nu_3 = \frac{5}{21}\ \tau_{\rm corr}\overline{u^{(0)2}}, \ \
 \nu_2 = \frac{32}{105}\ \tau_{\rm corr}\overline{u^{(0)2}}
 \label{16}
\ende
are always positive 

-- for white-noise
spectra: 
  \beg
  \nu_1 = \nu_2 = \frac{2\pi}{105\nu}\int\limits_{0}^{\infty}
  E(k)\frac{{\rm d}k}{k^2},\ \
  \nu_3 = \frac{\pi}{28\nu}\int\limits_{0}^{\infty}
  E(k)\frac{{\rm d}k}{k^2}
  \label{17}
 \ende
are
always positive. The 
  gap between (\ref{16}) and (\ref{17}) is filled by 
 \beg
 E(k,\omega ) = \frac{2}{\pi}\frac{w}{\omega^2 + w^2}\hat E(k). 
 \label{18}
 \ende
The coefficients (\ref{13}) are  positive-definite with  (\ref{18}). 

For small disturbances, $\tilde{\vec u}$, of the mean flow 
depending on $z$  the cross correlations defined by (\ref{5}) read
  \beg
  Q_{xz}&=&-\nu_0 D \tilde{u}_x + \nu_2 S D \tilde{u}_y,
  \nonumber\\
  Q_{yz}&=&-\nu_0 D \tilde{u}_y + \nu_3 S D \tilde{u}_x.
  \label{20}
 \ende
The linear equation system for the disturbances is
 \beg
  \frac{\partial \tilde{u}_x}{\partial t}&-& \nu_0 D^2 \tilde{u}_x
  + \nu_2 S D^2 \tilde{u}_y = 0,
  \nonumber\\
  \frac{\partial \tilde{u}_y}{\partial t}&+& S \tilde{u}_x
  - \nu_0 D^2 \tilde{u}_y + \nu_3 S D^2 \tilde{u}_x =0.
 \label{21}
 \ende
 It reduces to 
 \beg
 \left(\nu_2 \nu_3 S^2 - \nu_0^2\right) D^2 \tilde{u}_x
 + \nu_2 S^2 \tilde{u}_x =0
 \label{22}
 \ende
in the stationary case. The 1D problem should be accomplished by
boundary conditions imposed at (say) $z=0$ and $z = H$. 
But Eq. (\ref{22}) possesses a solution only if
 \beg
 H= \pi \sqrt{\frac{\nu_0^2 - \nu_2 \nu_3 S^2}{-\nu_2 S^2}}
 \label{23}
 \ende
irrespectively of  whether no-slip or stress-free boundary conditions are applied.
The second term in the numerator of (\ref{23}) must be small
compared to the first one. Otherwise the nonlinear correlations
(\ref{5}) which neglect the third and higher order terms in the
mean shear cannot be applied. Therefore, an instability  only exists for $\nu_2<0$. 
 The quantity $\nu_2$ proved to be positive for almost all spectra. The shear flow
(\ref{1}) thus proves to be stable in the hydrodynamic regime.

\end{document}